\documentclass[12pt]{article}
\usepackage{epsf,amssymb,amsmath}

\newcommand{\eq}[1]{\begin{equation} #1 \end{equation}}
\newcommand{\ar}[1]{\begin{eqnarray} #1 \end{eqnarray}}
\newcommand{\tr}{\mathop{\mathrm{tr}}\nolimits}

\def\ra{\rightarrow}
\def\const{{\rm const}}
\def\e{{\,\rm e}\,}
\def\d{\partial}
\def\D{\delta}

\newcommand{\br}[1]{\left( #1 \right)}
\newcommand{\vev}[1]{\left\langle #1 \right\rangle}
\newcommand{\rf}[1]{(\ref{#1})}
\newcommand{\non}{\nonumber \\*}
\def\const{{\rm const}}
\def\e{{\,\rm e}\,}
\def\d{\partial}
\def\D{\delta}

\def \vp {\varphi} 

\def \beq {\begin{equation}}
\def \eeq {\end{equation}}
\def \l {\lambda}

\def \s {\sigma}
\def \t {\tau}

\def \ci{\cite}

\def\lb{\lambda}
\def\st{\sqrt{\lambda}}
\def \adss {{$AdS_5 \times S^5$}}
\def \ov {\over}
\def \vp{\varphi}
\def \n {\nu}
\def \ep {\varepsilon}
\def \N {{\cal N}}

\def \vss {\psi}
\def \vff {\vp}
\def \vs0 {\psi_0}
\def \W {{\rm W}} 
\def \G {{\rm G}}

\begin{document}
\begin{titlepage}
\begin{flushright}
ITEP--TH--34/02
\end{flushright}
\vspace{.5cm}

\begin{center}
{\LARGE 
 Wilson loops in  $\cal N$=4 SYM theory:} 
\\[.25cm]
{\LARGE   rotation in $S^5$}\\
\vspace{.9cm} {\large A.A. 
Tseytlin $^{a,b,}$\footnote{
Also at Lebedev Physics Institute, Moscow, Russia. e-mail: tseytlin.1@osu.edu, 
tseytlin@ic.ac.uk.} 
 and
K. Zarembo $^{c,}$\footnote{Also
at ITEP, B. Cheremushkinskaja, 25, 117259 Moscow, Russia.
e-mail: Konstantin.Zarembo@teorfys.uu.se }}\\
\vspace{24pt}
{\it $^a$ Blackett Laboratory, Imperial College}
\\{\it London SW7 2BZ, U.K.}
\\ \vskip .2 cm
{\it $^b$ Smith Laboratory, The Ohio State University}\\
{\it Columbus, OH43210-1106, USA}
\\ \vskip .2 cm
{\it $^c$ Department of Theoretical Physics}\\
{\it Uppsala University, Box 803, S-751 08 Uppsala, Sweden}
\end{center}

\vskip .25cm
\begin{abstract} 
\noindent
We study  Wilson loops in $\N=4$   SYM theory 
which are non-constant in the scalar  ($S^5$) directions
and open string solutions associated with them
in the context of AdS/CFT correspondence. 
An interplay between Minkowskian and Euclidean pictures
turns out to be non-trivial for time-dependent Wilson loops. 
We find that in the 
$S^5$-rotating case 
there appears to be  no direct open-string duals
 for the  Minkowskian Wilson loops,
and their expectation values 
should be obtained by analytic continuation from  the 
Euclidean-space results.  In the Euclidean case, 
we determine the dependence of the ``quark -- anti-quark'' potential 
on the rotation parameter $\nu$, 
both at weak coupling (i.e. 
in the 1-loop 
perturbative  SYM theory) and at strong coupling (i.e. 
in the classical string 
theory in $AdS_5\times S^5$).

\end{abstract}
\end{titlepage}


\setcounter{page}{2}

\setcounter{footnote}{0}

\section{Introduction}

In the standard  picture \ci{review} of the AdS/CFT 
correspondence 
the closed string states in the bulk of $AdS_5$ 
are dual (at large $N$) to  the (single-trace)  operators 
in the $\N=4$ SYM  CFT.   
Recently, there was  a progress in understanding 
details of this correspondence in certain sectors  of string 
states with large 
(angular momentum) quantum numbers \cite{BMN,GKP}.

Adding a  D3-brane probe located parallel (and close) to the boundary introduces
open string sectors -- open strings attached to the D3-brane.
These are important  for the  description of  Wilson loops
as  basic non-local 
observables in the context
of AdS/CFT (for a review see  \ci{SZ}).

By analogy with  the closed-string case of \ci{BMN}, 
one  may wonder  if  considering special open string configurations 
with non-trivial dependence on  angles of $S^5$ may lead to
simplifications and  new interesting tests of the AdS/CFT.
It is natural to associate such open 
strings with Wilson loops that depend on scalar fields
in a non-trivial way.  
Here we shall study the dependence of the Wilson loops 
on a new parameter $\nu$  corresponding to rotation  of the open string 
configuration 
along big circle of $S^5$.  While  the corresponding Wilson loop 
will not carry a  definite R-charge and thus  there will be no 
direct relation to the closed string case of \ci{BMN}, 
we are motivated by 
the hope that an  investigation of more general 
scalar Wilson loops  
may  be useful for better understanding of  
 the AdS/CFT duality. In particular, one may be able to 
 identify new cases  where 
 certain features of SYM observables
can be interpolated from weak to strong 't Hooft coupling.

The {\it Minkowski}-signature 
Wilson loops  depending not only on a contour $x^m(\tau)$ in 
Minkowski space $\mathbb{R}^{1,3}$ 
but also on  a contour $\theta^i(\tau)$ 
in $S^5$  can be related to a propagator of a 
`` W-boson'' \ci{MAL,Rey,DGO} 
present   in the case of  a 
generic 
slowly varying scalar field condensate
(with $\theta_i(\tau)$ being  a unit 6-vector in the direction of the
symmetry breaking   condensate)
\eq{\label{wil} 
W(C,\theta)=\frac{1}{N}\,\tr{\rm P}\exp\left[i \oint_C
d\tau\,\br{ A_m(x)\dot{x}^m - \Phi_i(x)\theta^i|\dot{x}|}\right].
} 
Similar generalizations of constant scalar Wilson 
 loops ($\theta_i=\const$) were considered early on 
in the context of AdS/CFT correspondence \cite{MAL} 
 and were recently shown \ci{KZ} to admit new interesting 
cases in which supersymmetry is partially preserved.

Below we shall study  the simplest non-trivial case 
of ``rotation'' in $S^5$ when the unit 6-vector $\theta_i(\tau)$ is chosen as 
\eq{\label{coss} 
\theta^i= (\cos \nu \t, \sin \nu \t, 0,0,0,0) \, \ \ \ \ \ \ \nu=\const \ . 
}
In  the case of a static  source, i.e. single straight line
$x^m= (\t, 0, 0, 0 )$, that gives 
\eq{\label{wils} 
W= 
\frac{1}{N}\,\tr{\rm P}\exp\left[
i\int
d\tau\,\br{ A_0 (\t,{\bf 0})  - \Phi_1 (\t,{\bf 0})\ \cos \nu \t
-  \Phi_2 (\t,{\bf 0}) \ \sin \nu \t } 
\right] \ .   } 
The expression for the  corresponding  Wilson loop in the 
{\it Euclidean}-signature  case 
can  obtained by  the standard continuation  
$A_0 \to - i A_0,  \ \ x^0 \to i x^0, \   |\dot{x}|\equiv \sqrt{ - \dot x^2}  \to 
i  |\dot{x}| $, giving \ci{MAL,DGO}
\eq{\label{wilq} 
W(C,\theta)=\frac{1}{N}\,\tr{\rm P}\exp\left[ \oint_C
d\tau\,\br{i A_m(x)\dot{x}^m + \Phi_i(x)\theta^i|\dot{x}|}\right].
} 
In addition, it seems natural   to continue 
$\tau \to i \tau$. While this does not  affect the final result \rf{wil} 
for a {\it constant} $\theta_i$,  this  
may produce a complex expression  when $\theta^i$ 
is non-trivial   and some 
additional continuation  of parameters  may be needed.
For instance, if we change  $\tau$ to $i \tau$ in the Wilson loop
\rf{wils}, we should also substitute $-i\nu$ for $\nu$. The Euclidean
counterpart of the Wilson line with rotation in $S^5$ is 
\eq{\label{WLP}
W=\frac{1}{N}\,\tr{\rm P}\exp \left[\int d\tau\,
\br{iA_0(\tau,{\bf 0})+Z(\tau,{\bf 0})\e^{-i\nu\tau}
+\bar{Z}(\tau,{\bf 0})\e^{i\nu\tau}} \right]\ , 
}
where  we introduced  the complex scalar field 
$Z= {1 \ov 2} ( \Phi_1 + i \Phi_2)$. 

Our aim will be to compare the weak-coupling (i.e. perturbative SYM) 
and strong-coupling (i.e. semiclassical string theory in \adss)
predictions  for the expectation  values
of  such  Wilson loops, generalizing the previously known 
results for the single straight line and the antiparallel lines
\ci{MAL} to the case of $\nu\not=0$. 

While in the $\nu=0$ case the expressions for the 
Minkowski and Euclidean signature loops 
are directly related, we shall find that the two are  different 
in the rotating case. While on  the AdS side the Minkowski case  may 
seem natural  having a direct open-string energy interpretation, 
the corresponding perturbative SYM potential  turns out to be imaginary
(Section 2.1).
The latter is found to be a direct analytic continuation 
$\nu \to i \nu$  of the well-defined 
real  potential extracted from 
the  Euclidean Wilson loop  expectation value  (Section 3.1).
This  may be considered as an indication  that 
non-constant scalar Wilson loops should be associated 
with Euclidean minimal surfaces on the string side.


For $\nu=0$,
 the  simplest  1/2 supersymmetric  configuration (``straight line'')
corresponds to a  single W-boson at rest, i.e. to the 
$x^0=\tau$ line at the boundary of $AdS_5$
 which may be thought of as the end-line
of the open string stretched from the horizon $z=\infty$ to the boundary
$z=0$  \ci{MAL,Rey}.
  Below we shall first  determine how this  simplest 
classical open-string solution is modified 
in the rotating case.
 In the Minkowski case  (Section 2.2)
we  shall find  that instead of running all the way  to the horizon, 
the string reaches some maximal  height $z_0$ and then returns 
back to the same point at the boundary. 
In the Euclidean case (Section 3.2) the classical string solution 
is  still  similar to the $\nu=0$ case: 
 the infinite line along the radial AdS direction.
A novel feature is an instability, which leads to
shrinking of the minimal surface in  $S^5$.
 
We shall then generalize   the  minimal surface solution 
of \ci{MAL}  describing the  potential between two W-bosons
(i.e. 
the minimal surface 
ending on two  
anti-parallel lines on the boundary separated by distance $L$)
 to  the case with an extra 
``$S^5$-rotation'' characterized by the parameter $\nu$. In general,  
the potential will have the structure 
\eq{\label{pote}
V={1 \ov L} {\cal V} (l, \lambda) \ ,  \ \ \ \ \ \ l \equiv \nu L \ ,   }
i.e.  will be a non-trivial function  of the dimensionless
parameter $l= \nu L$   and the  't Hooft coupling 
$\lambda = g^2_{\rm YM} N = 
{R^4\ov \alpha'}$ (we assume $N \to \infty$).
At weak coupling, i.e. in the SYM 
perturbation theory 
\eq{\label{potqe}
V ={ \lambda \ov L} \G (l) + O(\lambda^2)  \ . }
At strong coupling, i.e. 
in the semiclassical string theory in \adss one finds that 
\eq{\label{pott}
V={\sqrt{\lambda} \ov L} \W(l)  + O(1)\  .}
The functions $\G(l)$ and $\W(l)$ will be determined below. 

We shall find 
that in the Minkowski case (Section 2.2)  the classical string solution 
no longer  exists if $l$ is larger than  certain critical  value 
$l_{max} \approx 0.678$.

In the Euclidean case (Section 3.2),  the minimal 
surface  exists for any  value 
of $l=\nu L$  but there  is a phase transition at $l=l_c
\approx 2.31$, which is similar
to the Gross-Ooguri transition 
\cite{GO,Zarembo:1999bu,Olesen:2000ji,Kim:2001td,Zarembo:2001jp},  
though the mechanism of the 
transition is different. In our case,
the phase transition happens due to the onset of an instability
for the minimal surface rotating in $S^5$, while the 
Gross-Ooguri phase transition 
is caused by the string breaking.
The potential is a non-analytic function of the distance, but
it has the same Coulomb-law ($1\ov L$) 
 asymptotics both at 
short and at large distances with the same effective charge.

\section{Minkowski Wilson loops with $S^5$-rotation}

\subsection{Perturbative  SYM  results 
}

\subsubsection{Single line}

The expectation value of the Wilson loop \rf{wils}  
is given, to 
the first order in perturbation theory in $\l$, 
by:\footnote{We use the metric with signature $(-+++)$
and the standard pole prescription for propagators.}
\eq{
\ln\vev{W(C_T)}=-\frac{\lb T}{2}\int_0^\infty d\tau\,
(1-\cos\nu\tau)\int\frac{d^4p}{(2\pi)^4}\,
\frac{-i\e^{-ip_0\tau}}{-p^2+i\ep}\,.
}
Here, $T$ is the length of the loop  
($-T/2<x^0=\tau< T/2$), which serves as an IR cutoff.
Computing the integrals over $p_0$ and over $\tau$, we get:
\eq{
\ln\vev{W(C_T)}=\frac{\lb T}{8\pi^2i}
\int_0^\infty dE\,E\br{\frac{1}{E}-\frac12\,\frac{1}{E+\nu}
-\frac12\,\frac{1}{E-\nu-i\ep}}.
} 
Now,
$$
\frac{1}{E-\nu-i\ep}=\wp\,\frac{1}{E-\nu}+i\pi\D(E-\nu)
$$
and
$$
\wp\int_0^\infty dE\,E\br{\frac{1}{E}-\frac12\,\frac{1}{E+\nu}
-\frac12\,\frac{1}{E-\nu}}=0.
$$
The only non-zero contribution comes from the pole at 
$E=\nu+i\ep$:
\eq{\label{doo}
\vev{W(C_T)}=\exp\br{-\frac{\lb}{16\pi}\nu T + O(\l^2)} \ .
}
This 
result  can be obtained from the  
Euclidean expectation value found in Section 3.1 by the substitution
$T\ra iT$, $\nu\ra -i\nu$. 

Usually, the exponent in the 
Wilson line vev is interpreted as a response of the
vacuum energy to the insertion of an external source. 
One would then expect an oscillating behavior
of the expectation value: $\vev{W(C_T)}=\exp(-iE_{\rm ext}T)$,
where $E_{\rm ext}$ is the energy which the source acquires due to
interactions with the Yang-Mills fields. Instead of oscillations,
we found 
an exponential decrease,
which, according to the standard lore, is a signal of   an instability.
One way to understand this instability is to recall that 
Wilson
line with time dependence in $S^5$ describes an infinitely heavy 
W-boson associated with
non-constant Higgs condensate. Time dependence of the condensate
allows the Wilson loop 
to emit on-shell scalars. As a result, the pole of the scalar 
propagator lies in the region of integration in the one-loop
amplitude, and the amplitude acquires
an imaginary part from the residue.

\subsubsection{Anti-parallel lines}

The potential $V(L,\nu)$ is defined  by the expectation
value of the rectangular loop whose time extent is much larger than
the spatial separation ($T\gg L$): 
\eq{
\vev{W(C_{T\times L})}=\e^{-iV T} \  .
} 
To the first order in perturbation  theory:
\eq{
V=-\frac{i\lb}{2}\int_{-\infty}^{+\infty}
d\tau\,(1+\cos\nu\tau)\int\frac{d^4p}{(2\pi)^4}\,
\frac{i\e^{-ip_0\tau+i{\bf pL}}}{-p^2+i\ep}\,.
}
After angular and energy integrations  we get 
\eq{
V =-\frac{\lb}{4\pi^2L}\int_0^\infty
dE\,\sin (EL) \,\br{\frac{1}{E}-\frac12\,\frac{1}{E+\nu}
-\frac12\,\frac{1}{E-\nu-i\ep}}.
}
Again, there is a resonant term and the potential picks
up an imaginary part from the pole at $E=\nu+i\ep$:
\eq{\label{daa}
V=-\frac{\l}{8\pi L} ( 1+\e^{i\nu L})   +  O(\l^2) \ .
}
This expression also turns out to be   an analytic continuation in $\nu$ of the
corresponding  Euclidean
potential found in Section 3.1.
This suggests   that  in general the expectation values of the Minkowski 
Wilson loops  can be obtained by an 
 analytic continuation from the  Euclidean
space.

As follows from the above expression, 
 the continuation in the leading-order perturbative 
expression  is non-trivial and
produces imaginary parts in the expectation values.
Almost certainly, continuation from
Euclidean space should  work to all orders of perturbation theory,
though it would be interesting to check this
by pushing perturbative
calculations to higher orders as it was done for the 
Wilson loops which are static in $S^5$ 
\cite{Plefka:2001bu,Arutyunov:2001hs}.
A natural guess then is that 
well-defined Euclidean Wilson loops are associated with
the Euclidean minimal surfaces at strong coupling.

Before turning to  the Euclidean case in Section 3, 
in the remainder of this section
we shall  consider  
 the Minkowski-signature  open strings
which end at the boundary of $AdS_5$. 
It seems that these string solutions are not directly
related to Wilson loops, but may be interesting on their own.


\subsection{Open strings in $AdS_5$  with Minkowski 
signature
 }

We shall use the \adss{} metric in the following form
(Poincare coordinates) 
\eq{\label{hjk} 
ds^2= {1 \ov z^2}
 (dx^m dx^m + dz^p dz^p  ) \ , \ \ \ \
 z^2 = z^p z^p \ , \ \ \ p=1,...,6 
 } 
\eq{\label{hqj}  dx^m dx^m = - dx^0 dx^0 + dx^k dx^k \ , \ \  \ \ \ \ 
k=1,2,3
}
Let us  first recall the form of the open string solutions 
in the absence of  rotation, i.e. for  $\nu=0$. We shall use the conformal gauge 
on the world sheet. 
In the case of single  straight line on the boundary  along $x^0$, 
the string is stretched along the radial AdS direction
 (here we use Minkowski signature  in 
both target space and string world sheet)
\eq{\label{hj}
x^0= \t \ , \ \  \ \ \ \ \ \ \    z= \s \ ,  \ \ \ \  \  \ \ \ \ \ \ 
 0 < \s < \infty 
\ .   
 }
The Minkowski version  of the two anti-parallel lines 
configuration \ci{MAL}   is 
(prime is derivative over $\s$)  
\eq{\label{hjos}
  x^0 =\t \ , \ \ \ \  x^1=x^1(\s)\equiv  x(\s) \ , \ \  \ \ \ z=z(\s) \ ,  
 }
\eq{\label{hji}
x' = c z^2 \ , \ \ \  \ \ \  {z'}^2 = 1 - c^2 z^4 \ , \ \ \  \ \ 
z_{max} = { 1 \ov \sqrt{c}} \ , \ \ \ \ \ \ \ 
\br{{dz\over dx}}^2  =( c^2 z^4)^{-1}  -1. 
}
$c=0$ is the  straight line case. 
Both solutions admit 
straightforward Euclidean  analogs   obtained by 
$ \t \to i \t \ , \  \  x^0 \to i x^0 .$

\subsubsection{Single line}

A simple generalization  of  the straight line solution \rf{hj}
is obtained
by adding a  rotation along the  big circle of $S^5$,
i.e. in the 2-plane  ($z_1,z_2)$ in  the 6-space
\eq{\label{whj}
x^0 = \t\ , \ \ \ \  \vp =\nu \t \ , \ \ \  \  z=z(\s) \  , 
\ \ \ \ \ \ \
z_1 + i z_2 = z e^{i \vp} \ ,  }
\eq{\label{hjo}
 {z'}^2 = 1  - \n^2 z^2 \ ,\ \ \ 
{\rm i.e. } \ \  \  \   
\ z= {\nu}^{-1}  \sin \n \s  \ .  }
For zero  angular momentum parameter 
$\n\to 0$ we get  back  to the straight line solution \rf{hj}.
 For $\nu\not=0$  the radial coordinate
 $z$ is periodic in $\s$:  for fixed $\t$ the string starts 
 at the boundary, reaches the  maximal value  $z_{max} = {1  \ov \nu}$
  and then returns back to the same point $x_k=0$ at the
 boundary. For changing $\t$  it   rotates  in the  $\vp$-direction  in $S^5$
with speed $\nu$. 
As we shall see below, 
this  configuration may be thought of as a limit of a string ending 
at the two points  of the boundary  in the case 
when these points coincide.
One  may restrict $ \s$ to run from $0$ to $ \pi\ov \nu$, 
so that the string  has only one fold
(multi-folded strings will have bigger energy).

This open-string configuration is characterized by the  values of the conserved 
charges -- the  space-time energy and the angular  momentum
\eq{\label{whjs}
E= 2\sqrt \l \int^{{\pi\ov 2\nu} }_0  {  d \s \ov 2 \pi}  z^{-2}  \dot x^0   \ , \ \ \ \ \  
 J = 2\sqrt \l \int^{{\pi\ov 2\nu} }_0 {  d \s \ov 2 \pi} \dot \vp    \ , 
 }
where we added factors of 2 as the integral goes from 0 to the folding point.
Not 
surprisingly, the value of the angular momentum does not depend on $\nu$
($J$ is dimensionless, $\nu$ has a dimension of mass):
\eq{\label{qhj}
 J= {  \sqrt \l \ov 2 }   \ . } 
The expression for the energy is 
divergent at $\s=0$ and can be defined  by subtracting,
 as in the $\nu=0$ case, the value
of the infinite straight string configuration. Note that we have to subtract the  straight-line contribution twice 
as the two segments of the folded string  include the singular point $z=0$.
Then  the subtracted value of the energy is 
\eq{\label{whkj}
E= 
 2 \sqrt \l \br{\int^{\pi\ov 2\nu}_0  {  d \s \ov 2 \pi} 
  {\nu^2 \ov \sin^2\nu \s }  -  
 \int^{\infty}_0  {  d \s \ov 2 \pi\s^2 }
}   =  
  { \sqrt \l \ov \pi} 
 \left.\left( {\nu \cot \nu \s } - { 1  \ov   \s } \right)
\right|_{\s=0 }
  = 0 \ . }  
Thus the subtracted energy vanishes as in the non-rotating case, 
despite the fact that the $\nu\not=0$ solution is no longer supersymmetric.
\footnote{Note that in the 
the Euclidean signature  case  in Section  3.2
 the subtracted  value of the Euclidean action  will turn out to be non-vanishing.}

One may wonder if the solution  with $\tau$-dependent angle of $S^5$ 
and no other angles excited is actually stable. 
This is indeed so  in Minkowski  signature.
Including the two-sphere directions $\psi$ and $\vp$ 
of $S^5$ 
the relevant 
part of the string action in the conformal  gauge is 
\eq{\label{whokj}
S ={ \sqrt \l \ov 4 \pi}   \int  {  d \s d \tau }
\left[  {1 \ov z^2} ( \dot x_m^2  + \dot z^2 -  {x'}_m^2  + {z'}^2  )
   + \dot  \psi^2  - {\psi'}^2    + 
   \cos^2 \psi\  ( \dot  \vp^2  - {\vp'}^2) \right]. 
   }
For $\vp= \nu \tau$ the effective potential for $\psi$ is 
${\cal U}= {\psi'}^2   - \nu^2 \cos^2 \psi $.   Its minimum  ${\cal U}
=-\nu^2$ is indeed at 
$\psi= 0$, as we were assuming above. 
The stability of this Minkowski
 solution can be  also confirmed by the direct analysis 
of the action for  small fluctuations as in \ci{DGT,Frolov:2002av}.


The Euclidean analog of the  above solution  with $\psi=0$ 
is obtained by  the following 
continuation:
$\t \to i \t, \ x^0 \to i x^0, \  \nu \to -i \nu$, 
i.e. 
\eq{\label{ophj} 
x^0 =  \t\ , \ \ \  \vp =\nu \t \ , \ \ \ 
 {z'}^2 = 1 + \n^2 z^2 \ ,  \ \ {\rm i.e.} \ \ \ 
  \ z= {1 \ov \nu} \sinh \n \s  \ .  }
Here we get again an infinite open string: $\s$ and thus $z$ 
changes from 0  to infinity. 
\label{instab} This Euclidean solution is, however, 
 unstable, because its area linearly diverges
at the horizon. Its  stable counterpart involves 
non-constant $\psi$-angle  and will be described in Section 3.2.

\subsubsection{Anti-parallel  lines}

Let  us now consider the $\nu\not=0$ generalization of the 
two-line solution \rf{hjos}.\footnote{This generalization was also 
   independently  studied  by J. Russo.}
This will be 
  an open string with both ends at the boundary separated  by distance $L$, 
stretched in $z$
direction  and rotating  along the angle $\vp$  in $S^5$.
We shall set the angle $\psi=0$. 
This solution is similar to the 
Euclidean solution A in Section  3.2.

For $\vp=\nu\tau$   and $x_1= x(\s)$    one finds 
(cf. \rf{hji})
\eq{\label{okj}
  x' = c \  z^2 \ , \ \  \ \ \ \ 
\ \ 
{z'}^2 = 1- \nu^2 z^2  - c^2 z^4, 
}
i.e. \eq{\label{ej}
{z'}^2 =  (1- b^2 z^2) [ 1 +  (b^2-\nu^2)  z^2]  \ , \ \ \ \
 \ \ \  \ c= b \sqrt{ b^2-\nu^2 }.  } 
We introduced, instead of the integration constant $c$,  
the parameter   $b \geq \nu$, so that   $b^{-1}$ is the maximal
value of $z$. 
Then 
\eq{\label{ikj}
 L=\int dx   =  2 {\sqrt{1 - v^2} \ov b} 
\int^1_0 {w^2 dw   \over \sqrt{(1-w^2) [ 1 + (1-v^2) w^2]} } \ ,}
\eq{ 
w= b z \ , \ \    \ \ \ \ \ \ \   v\equiv  { \nu \ov b } \   .  } 
Here $0 \leq v \leq 1$. 
In the   case of $\nu=0$  (i.e. $v=0$)
\eq{\label{iikj}
 L= \int dx = b^{-1}  k_0 \ , \ \ \  \ \ \ \ \ 
 k_0 =  {(2 \pi)^{3/2} \over [\Gamma({1\over 4})]^2 }\approx   1.198  \ . } 
Eq. \rf{ikj} 
determines  the maximal value $z_{max}= b^{-1}$ in terms of $L$ and $\nu$. 
Equivalently, it gives  a  relation  between the two dimensionless 
parameters  $l= \nu L$ and $v =  { \nu \ov b } $:
\eq{\label{ipkj}
l =  2 { v  \sqrt{1 -v^2} }   
\int^1_0 {w^2 dw   \over \sqrt{(1-w^2) [1 +  (1-v^2) w^2] } } \ .  }
The dimensionless distance, as a function of $v$,\ 
$l(v)$, vanishes at $v=0,1$  and, consequently,  has maximum  at
some $0<v_0<1$ (numerically, $v_0\approx 0.75$).
As a result, 
the string solution is possible  only for $l=L \nu$ smaller than
$l_{max}=l(v_0)\approx 0.678$. For fixed  $\nu$ we cannot make  $L$ arbitrarily large, 
or, equivalently, for fixed $L$ there exists a maximal value of $\nu$ 
above which the solution  does not exist. 
As we change the parameter $v$ from zero to one, $l$ first grows,
then reaches its maximal value and then returns back to 
zero. To summarize, the 
 solution cannot be continued past the maximal distance between
the lines, and the classical string world sheet cannot connect
lines on the boundary which are separated by a larger distance.
 
\begin{figure}[h]
\begin{center}
\epsfxsize=9cm
\epsfbox{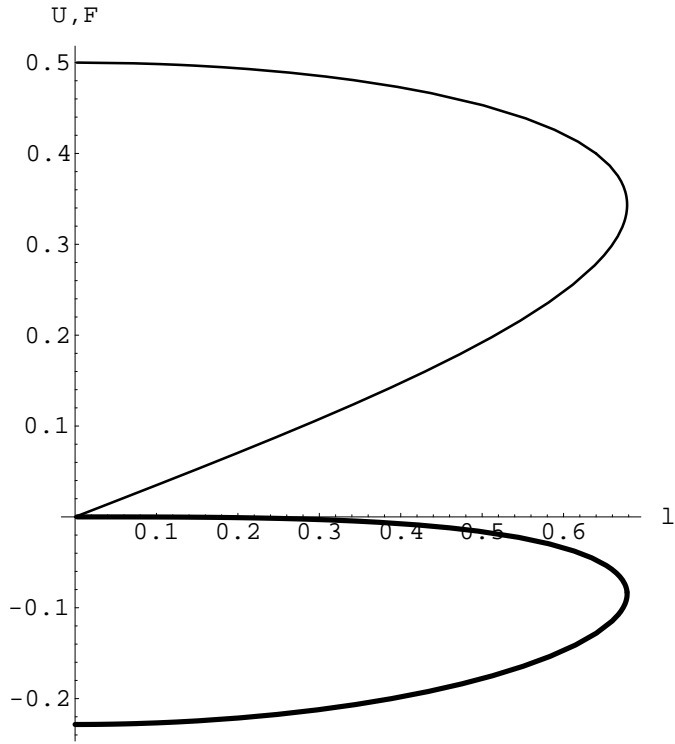}
\end{center}
\caption[x]{\small The energy  ($U(l)$, bold line)
and the angular momentum ($F(l)$) as functions of $l=\nu L$.
Both curves go from  bottom to top as $v$ goes from
0 to 1.}
\label{fig2}
\end{figure}

The energy and the angular momentum of the string are  given by 
(subtracting an infinite
constant from the energy): 
\ar{
E  &=&2 { \sqrt \l \ov  2\pi}    \int {d  \s  \ov  z^2}  - E_0 
\non
  &=& {\sqrt \l \ov  \pi L } {l  \ov  v    }   \left\{ \int^1_0  {d w \ov w^2}  
 \left[\frac{1}{\sqrt{(1-w^2)[ 1 +  (1-v^2) w^2] }}-1 \right]
-1\right\}
\non
&=&-\frac{\sqrt{\l}}{L}\,\frac{l^2\sqrt{1-v^2}}{2\pi v^2}\non
 &\equiv&    { \sqrt \l    \ov L}\, U(l), 
\\*
 J &=& 2 { \nu  \sqrt \l \ov  2\pi}    \int { d \s } 
 =  { \nu  \sqrt \l \ov  \pi}
  \s_{max}(v) =  
 {   \sqrt \l \ov  \pi  }  { v  } 
    \int^1_0 { d w\ov  \sqrt{(1-w^2) [1 +  (1-v^2) w^2]  } }  
\non
&\equiv& \sqrt \l\,  F (l).
} 
The functions $U$ and $F$ determining the dependence of $E$ and $J$ on $\nu$ 
 are shown in fig.~\ref{fig2}.\footnote{We use the notation $U$ instead of 
$\W$ in \rf{pott} to distinguish the Minkowski case from the Euclidean one.}
 It should be noted that very similar
expressions for the energy and the angular momentum  arise also for 
a time-independent
string solution, where the string  is stretched along the big circle of
$S^5$ \cite{MAL}.

The angular momentum $J$  
grows from $0$ at $v=0$ to its  maximal value ${\sqrt \l \ov 2}$ at $ v=1$, 
which corresponds to the 
folded single-line solution
discussed above. 
The potential vanishes at the point where $J$ has its maximum
value.
In contrast to the closed string case in \ci{BMN}, here the angular momentum 
$J$ does not have an intrinsic  meaning 
on the gauge-theory side 
so it does not seem natural to express $E$ as a function of $J$.

We conclude that there is no similarity between  the perturbative 
SYM  and the semiclassical string-theory expressions   
for the Minkowski-signature 
Wilson loop  expressions 
 in the $\nu\not=0$ case. 
We shall therefore  turn  now  to the study of the Euclidean signature case
where the results  will be more consistent with the AdS/CFT duality. 

\pagebreak

\section{Euclidean Wilson loops with rotation in $S^5$}

\subsection{Perturbative  SYM  results 
}

\subsubsection{Single line}

We start with  the  Euclidean 
 Wilson line \rf{WLP}  which  rotates 
 in $S^5$.  
The expectation value of this operator is an even function of
the rotation parameter $\nu$, which has the dimension of mass.
Actually, $\nu$  is the only dimensionful parameter in the
problem, except for an IR cutoff $T$.
On dimensional grounds, $\nu$ must enter in the combination $|\nu|T$.
We expect the expectation value to have an exponential form:
\eq{\label{gyg}
\vev{W(C)}=\e^{-f(\lambda)|\nu| T}~~~~~{\rm for }~T\rightarrow\infty\  ,
}
where $f(\lambda)$ is some function of the 't~Hooft coupling.
Because the expectation value depends on $|\nu|$, it
 is not analytic in $\nu$ at $\nu\rightarrow 0$. This turns out to
be a generic feature of Euclidean Wilson loops which are non-constant 
in  $S^5$. 
Below, we assume that $\nu>0$ and thus write $\nu$ instead of $|\nu|$.

A simple one-loop calculation gives (cf. \rf{doo}) 
\eq{
f(\lambda)=\frac{\lb}{8\pi^2}\int_{0}^{+\infty}
d\tau\,\frac{1-\cos(\nu\tau)}{\tau^2} + O(\l^2) 
=\frac{\lb}{16\pi} + O(\l^2) \ . 
}

\subsubsection{Antiparallel lines}

Next, let us  consider the potential between two heavy sources, each of
which involves rotation in $S^5$ with the same frequency. This
corresponds to the anti-parallel Wilson lines separated by the 
distance $L$.
To determine the potential, we should subtract the self-energy:
\eq{\label{kii}
\frac{\vev{W(C_{L\times T})}}{\vev{W(C_T)}^2}
=\e^{-V(L)T}~~~~~~~~{\rm for~~~}T\gg L\ .
}
At weak coupling one finds (cf. \rf{daa})
\eq{\label{duu} 
V(L)=-\frac{\lb}{8\pi^2}\int_{-\infty}^{+\infty}
d\tau\,\frac{1+\cos(\nu\tau)}{\tau^2+L^2} + O(\l^2) 
=-\frac{\lb}{8\pi}\,\frac{1+\e^{-\nu L}}{L} + O(\l^2) \ .
}
This determines the function $\G(l)$ in \rf{potqe}.

\subsection{Open string minimal surface 
 }

\subsubsection{Single line}

The AdS dual of the Wilson loop \rf{WLP} is a straight string which 
 starts  near the boundary of $AdS_5$  and rotates in $S^5$.
 Since  the rotation at infinity 
leads to linearly divergent area, the string will
slide down the sphere as it moves towards  the interior of $AdS_5$.
Such a solution corresponds to the ansatz
\eq{\label{old}
x^0=x^0(\tau),~~~~~\vff=\vff(\tau), ~~~~~z=z(\sigma),~~~~~\vss=\vss(\sigma)\ .
}
The string action is the Euclidean analog of \rf{potqe}
\eq{\label{act}
S=\frac{\sqrt \l}{4\pi}\int d^2\sigma\,\left[
\frac{(\d_a z)^2+(\d_a x^m)^2}{z^2}+(\d_a \vss)^2+\cos^2\vss\,(\d_a \vff)^2
\right].
}
The minimal-area solution can be found by satisfying the conformal-gauge constraint separately 
on the $AdS_5$ and $S^5$ parts. 
The $AdS_5$ part of the solution   is then the  same as for $\nu=0$:
\eq{
x^0=\tau,~~~~~~~~~~z=\sigma.
}
For the rotating string, we also have
\eq{
\vff=\nu\tau \ .
}
The remaining constraint (i.e. 
the equation of motion for the azimuthal angle $\psi$)  can 
then be easily solved to give:
\eq{
\sin\vss=\tanh (\nu\sigma) \ .
}
The metric induced on the minimal surface 
is that of the $AdS_2$+$S^2/\mathbb{Z}_2$:
\eq{
ds^2=\frac{d\tau^2+d\sigma^2}{\sigma^2}+
\nu^2\,\frac{d\tau^2+d\sigma^2}{\cosh^2(\nu\sigma)}\,.
}
The area of $AdS_2$ is subtracted by the regularization, 
while the area of  the hemi-sphere
is
\eq{\label{disc}
A=\nu T\  .
}
In general, the  single-line vev depends non-trivially only
on the 't~Hooft coupling, while its dependence on $\nu$ is determined by
the conformal symmetry.
Including the string tension factor, 
 the strong-coupling prediction for the function $f$ in \rf{gyg} is thus 
\eq{
f(\lb)=\frac{\st}{2\pi} + O( 1) \ . 
}
Hence, the function $f(\lambda)$ is expected to smoothly
interpolate between  $\sim\lambda$ at week coupling and 
$\sim\sqrt{\lambda}$ at strong coupling, similarly
to many other observables in ${\cal N}=4$  SYM theory. 

\subsubsection{Anti-parallel lines}
As was already mentioned above, 
on dimensional grounds, the potential 
between antiparallel lines is a function of 
$
l=\nu L, 
$
i.e. is given by \rf{pott} where the function $\W(l)$ 
is proportional to  the area per unit length
for the surface that
connects two anti-parallel lines, with
the area of the two disconnected surfaces \rf{disc} subtracted.

The ansatz for the minimal surface is a generalization of  \rf{old}
\eq{
x^0=\tau, ~~~~~\vff=\nu\tau,~~~~~~~
  z=z(\sigma),~~~~~x^1=x(\sigma),~~~~~
\vss=\vss(\sigma) \ . 
}
The action is the same as in \rf{act}.
The resulting equations  of motion can be integrated:
\ar{
x'&=&c\ \nu^2z^2,
\non
z'&=&\pm\sqrt{1+a\nu^2z^2-c^2\nu^4z^4},
\non
\vss'&=&\pm\nu\sqrt{\cos^2\vss-\cos^2\psi_0},
}
where $a$, $c$ and $\cos^2\psi_0$ are constants of
integration. 
The constraint,
\eq{
\frac{{z'}^2+{x'}^2-1}{z^2}+{\vss'}^2-\nu^2\cos^2\vss=0,
}
reduces to the condition
\eq{
a=\cos^2\psi_0.
}
It is convenient to introduce dimensionless variables 
$$\zeta=\nu z \, \ \ \ \ \ 
\xi=\nu x \ ,\ \ \ \ \
s=\nu\sigma \ . $$  They
satisfy the following equations:
\ar{\label{solut}
\xi'&=&c\zeta^2,
\non
\zeta'&=&\pm\sqrt{1+\cos^2\psi_0\zeta^2-c^2\zeta^4},
\non
\vss'&=&\pm\sqrt{\cos^2\vss-\cos^2\psi_0}.
}
The solution is symmetric under the  reflection
of $s$. The upper signs should be chosen
to the left of the turning point $s_0$
and the lower signs to the right. 
 The derivatives
of $\zeta$ and $\vss$ must vanish at $s=s_0$. Consequently,
$\vss(s_0)=\psi_0$. The AdS radius
at the turning point, $\zeta(s_0)=\zeta_0$, is
the root of the equation:
\eq{
1+\cos^2\psi_0\ \zeta_0^2-c^2\zeta^4_0=0.
} 
It is convenient to consider one half of the solution
to the left of the turning point and then continue it by
symmetry. 

There are two possible solutions which we shall call ({ A})  and ({ B}).
One of them does not depend on the
 azimuthal angle: $\psi\equiv 0$,
or $\psi_0=0$. The boundary conditions fix the remaining constant
of integration:
\eq{
\frac{l}{2}=c\int_0^{\zeta_0}
\frac{d\zeta \,\zeta^2}{\sqrt{1+\zeta^2-c^2\zeta^4}}\,,~~~~~
\psi_0=0~~~~~({\bf A})
}
In the other  solution, the azimuthal angle is non-trivial,  and the 
boundary conditions reduce to
\ar{\label{const1}
\frac{l}{2}&=&c\int_0^{\zeta_0}
\frac{d\zeta \,\zeta^2}{\sqrt{1+\cos^2\psi_0\zeta^2-c^2\zeta^4}}
~~~~~({\bf B})
\\* \label{const2}
\int_0^{\psi_0}\frac{d\psi}{\sqrt{\cos^2\psi-\cos^2\psi_0}}
&=&
\int_0^{\zeta_0}
\frac{d\zeta}{\sqrt{1+\cos^2\psi_0\zeta^2-c^2\zeta^4}}
~~~~~({\bf B})
}
The straightforward calculations of the induced metric and the area, i.e. 
the function $\W(l)$ in \rf{pott}, 
for the two solutions 
give:
\ar{\label{area1}
 \W(l)&=&2l\left[\int_0^{\zeta_0}
\frac{d\zeta}{\zeta^2}
\br{\frac{1}{\sqrt{1+\zeta^2-c^2\zeta^4}}-1}
-\frac{1}{\zeta_0}-1
\right. \non &&\left.
+\int_0^{\zeta_0}
\frac{d\zeta}{\sqrt{1+\zeta^2-c^2\zeta^4}}\right]
\non
&=&l\left[-cl-2+2\int_0^{\zeta_0}
\frac{d\zeta}{\sqrt{1+\zeta^2-c^2\zeta^4}}\right]
~~~~~({\bf A})
}
\ar{\label{area}
 \W(l)&=&2l\left[\int_0^{\zeta_0}
\frac{d\zeta}{\zeta^2}
\br{\frac{1}{\sqrt{1+\cos^2\psi_0\ \zeta^2-c^2\zeta^4}}-1}
-\frac{1}{\zeta_0}-1
\right.
\non
&&\left.\vphantom{\int_0^{\zeta_0}
\frac{d\zeta}{\zeta^2}
\br{\frac{1}{\sqrt{1+\cos^2\psi_0\ \zeta^2-c^2\zeta^4}}-1}
-\frac{1}{\zeta_0}-1}
+\int_0^{\psi_0}
\frac{d\psi\,\cos^2\psi}{\sqrt{\cos^2\psi-\cos^2\psi_0}}\right]
\non
&=&l\left[ -cl-2+2\int_0^{\psi_0}
\frac{d\psi\,\cos^2\psi}{\sqrt{\cos^2\psi-\cos^2\psi_0}}\right]
~~~~~({\bf B})
}
 Here we have 
subtracted  the area of $AdS_2$  ($\int {d\zeta\ov \zeta^2}$)
for the sake of regularization  as well as the  self-energy \rf{disc}
from the bare action calculated on the solutions of \rf{solut}.

\begin{figure}[h]
\begin{center}
\epsfxsize=9cm
\epsfbox{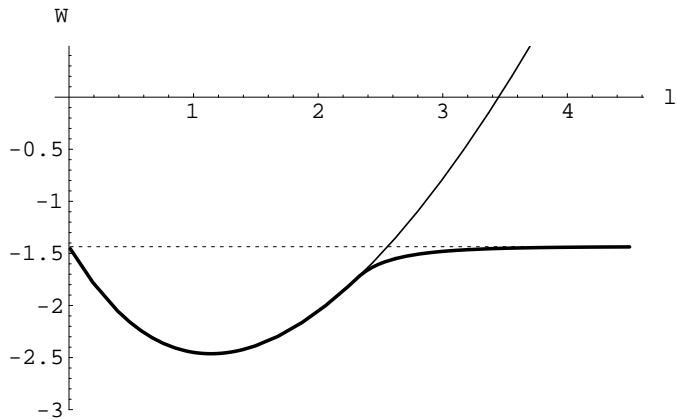}
\end{center}
\caption[x]{\small $W(l)$ as a function of $l$ (bold curve).
The area of the unstable solution A (thin curve). The dashed line
corresponds to pure Coulomb potential at $\nu=0$.}
\label{fig1}
\end{figure}

We have computed the  function $\W(l)$ in the potential \rf{pott} 
numerically.
The results are shown in fig.~\ref{fig1}.

\subsubsection{Phase transition}

The rotation in $S^5$ effectively gives mass to the azimuthal
fluctuations of the string: $(\d\varphi)^2=\nu^2$ in \rf{act}.
Since the mass-squared is negative, the solution with $\psi=0$ 
is clearly unstable. However, for a sufficiently short string,
this instability has no time to develop.

To see this, let us analyze the equation of motion for the azimuthal
angle:
\eq{
\frac{d^2\vss}{ds^2}+\cos\vss\,\sin\vss=0 \ ,\ \ \ \ \ \ s=\nu \s \ . 
}
This is the Newton's equation for a pendulum. As the string propagates
from one Wilson line to another, the azimuthal angle changes
from $\vss=0$ to $\vss=\psi_0$, and then bounces back to $\vss=0$,
so
the pendulum makes one half of
the oscillation. The total
period of oscillation is equal to $4s_0$,
twice the internal length of the string. The period depends on
the energy: it is bigger for larger energy. The total energy
is equal to the potential at the turning point, $-\cos^2\psi_0$.
Hence, the
smaller $\cos\psi_0$, the larger is the period. When $\cos\psi_0$
reaches zero, the period goes to infinity. But the period cannot
be arbitrarily small. It approaches a finite limit as $\cos\psi_0$ 
goes to 1. The smallest period is equal to the period of 
harmonic oscillations 
around the minimum
of the potential.
Consequently, sufficiently short string, whose length
is smaller than half-period of the harmonic oscillations, 
cannot move in the azimuthal
direction. 

So, the solution A with
$\cos\vss=1$
is stable for sufficiently small $l$, when the string 
is too short and the solution B does not exist.
At sufficiently large $l$, the string becomes
long enough to support the motion in the azimuthal direction
and then the non-trivial solution
takes over. As a consequence, the area and the potential are
continuous, but non-analytic functions of the distance.
There is a transition from one solution to another which
leads to non-analyticity of the potential.
In a similar context, the non-analyticity (Gross-Ooguri phase 
transition \cite{GO,Zarembo:1999bu,Olesen:2000ji,Kim:2001td,Zarembo:2001jp})
  occurs due to the string breaking.
Here the mechanism is different. The order of the transition 
is also different. In the present case, the potential undergoes, 
as will become clear shortly,
a second
order transition,  while  string breaking lead to the first order
transition.

The solution B ceases to exist at the critical
separation between the lines $l=l_c$, at which
the left-hand side of \rf{const2} reaches its minimal value $\pi \ov 2$: 
\eq{
\frac{\pi}{2}=\int_0^{\zeta_0}
\frac{d\zeta}{\sqrt{1+\zeta^2-c_c^2\zeta^4}}\,.
}
The first constraint \rf{const1}
then determines the critical distance
\eq{
\frac{l_c}{2}=c_c\int_0^{\zeta_0}
\frac{d\zeta
\,\zeta^2}{\sqrt{1+\zeta^2-c_c^2\zeta^4}}\,.
}
Numerically, $l_c\approx 2.31$. The areas of the solutions A and B are
equal at the critical point, so the transition is of the
second order here: the area is continuous together with its first
derivative. 

It is easy to analyze the potential in the two limiting cases of small 
and large  distances:

\bigskip

{\bf Short distances: $L\rightarrow 0$}
\bigskip

The area
\rf{area1} and the equation for $c$, \rf{const1},
differ from the 
results at $\nu=0$ \ci{MAL} by the last term in \rf{area1} and
by the extra
$\zeta^2$
in the argument of the square root. These differences
disappear
as $l\rightarrow 0$, since then $c\rightarrow\infty$
and the region
of integration with $\zeta\sim c^{-1/2}$ gives the dominant
contribution. To the leading order in $1/l$, the 
computation
of the potential repeats the  calculation of \ci{MAL}, so 
\eq{
V(L)=-\frac{4 \pi^2 }{[\Gamma(1/4)]^4}\cdot \frac{ \sqrt \lambda }{L} 
+\ldots~~~~~(L\rightarrow 0).
}

\bigskip

{\bf Long distances: $L\rightarrow\infty$}
\bigskip

The solution at large distances is very different, but,
surprisingly, we get the same potential, although
for a different 
reason: at $l\rightarrow\infty$ the length of the
string
becomes very large,  and the period of oscillation
in the azimuthal direction
must be large too. In other words, $\cos\psi_0$ rapidly 
converges to zero as $l$ goes to infinity. Then the  last
two  terms in \rf{area}  almost cancel each  other with
an exponentially
small leftover.
Also, the quadratic piece in $\zeta$ in the argument of
the square 
root gets multiplied by an exponentially small
factor
and can be neglected. What remains after such
approximations is
the same formulas as in the calculation of \ci{MAL}.
As a result, we get the same expression for
the potential, up to exponentially small corrections (cf. \rf{duu}):  
\eq{
V(L)= -\frac{4 \pi^2 }{[\Gamma(1/4)]^4}\cdot \frac{ \sqrt \lambda }{L} 
+\ldots~~~~~(L\rightarrow \infty) \ . 
}

\section{Discussion}

We have studied open string solutions which involve rotation in $S^5$ 
with a constant frequency. In the SYM theory,
these solutions are naturally associated
with Wilson loops whose coupling to scalars corresponds 
to rotation in a plane in internal 6-dimensional space.
Unexpectedly, we found that Minkowskian and Euclidean solutions
are absolutely different and cannot be obtained from each
other by  analytic continuation. 

On the other hand, 
for the  perturbative SYM expectation values
of the Wilson loops the analytic continuation does work,
though time dependence of the scalars ($S^5$ coordinates)
makes Wick rotation quite subtle. After continuation
from Euclidean to Minkowski space,  vev's of rotating Wilson loops
become complex-valued.
This  presents   a serious obstacle for 
their stringy
interpretation directly in Minkowski-signature $AdS_5 \times S^5$  space. 
It seems likely that  the duality between Wilson loops and open strings 
is well-defined only in the Euclidean version of the
AdS/CFT correspondence. 

The role  of the classical solutions for the open strings
 in the Minkowski case described in Section 2.2 
is unclear to us. It would be very interesting 
to clarify their meaning and
 to understand what do they correspond to on the gauge theory
side of the AdS/CFT duality. 

One of the lessons of this work 
is that  there seems to be  no direct relation between 
the open and closed
string states with  rotation  in $S^5$. This is probably
 related to the fact
there is no sense in which a Wilson loop
can  carry a definite  R-charge. Indeed, a  generic Wilson loop
is a mixture of states with various R-charges --  its 
local expansion contains 
 operators
of different R-charges \cite{Shifman1980:ui}. Perhaps, a study of
the OPE coefficients \cite{Berenstein:1999ij,Semenoff:2001xp} 
for operators with  large R-charge may help
to establish a contact between the semiclassical string picture
of \cite{BMN,GKP} and the Wilson loops.

\subsection*{Acknowledgments}
We are grateful to O.~Aharony and J.~Russo for useful discussions 
and to J.~Russo for
informing us about  his related unpublished   results. 
The work of A.~A.~T.  was  supported in part by
 the DOE grant
DE-FG02-91ER40690, the  PPARC grant SPG 00613,
 INTAS  99-1590 and the Royal Society  Wolfson
 research merit award.
The work of K.Z. was supported by the Royal Swedish Academy of Sciences
and by STINT grant IG 2001-062
and, in part, by RFBR grant 02-02-17260 and grant
00-15-96557 for the promotion of scientific schools.

\pagebreak

\end{document}